\documentstyle{article}

\newcommand{\be}{\begin{equation}}
\newcommand{\ee}{\end{equation}}
\newcommand{\bea}{\begin{eqnarray}}
\newcommand{\eea}{\end{eqnarray}}
\newcommand{\re}[1]{eq.(\ref{#1})}
\newcommand{\cri}[3]{\left\{\hspace{-.2cm}\begin{array}{c} #1\\ #2\end{array}
\hspace{-.2cm} \right\} }

\title{\bf Geodesics or autoparallels from a variational principle?} 
\author{Nuno Barros e S\'a\\
\\
\it Fysikum, Stockholms Universitet, Box 6730,\\
\it 113 85 Stockholm, Sverige\\
and\\
\it DCTD, Universidade dos A\c cores, R.M\~ae de Deus,\\
\it9500 Ponta Delgada, A\c cores, Portugal\\
\\
nunosa@vanosf.physto.se\\
\\}

\begin{document}

\vbox{USITP 97-19

\maketitle}

\begin{abstract}

Recently it has been argued that autoparallels should be the correct
description of free particle motion in spaces with torsion, and that such
trajectories can be derived from variational principles if these are suitably
adapted. The purpose of this letter is to call attention to the problems
that such attempts raise, namely the requirement of a more elaborate
structure in order to formulate the variational principle and the lack of a
Hamiltonian description for the autoparallel motion. Here is also raised the
problem of how to generalize this proposed new principle to quantum mechanics
and to field theory. Since all applications known of such a principle are
equally well described in terms of geodesics in non-holonomic frames we
conclude that there is no reason to modify the conventional variational
principle that leads to geodesics.

\end{abstract}

\bigskip


\newpage

\section{Introduction}

The motion of a free particle is a fundamental problem in physics. It is the
ultimate manifestation of the principle of inertia, that has been formulated
in different manners throughout the history of physics, and which must be
expressed somehow in any good theory. In quantum theory, for example, where
point particles are not fundamental objects, it must nevertheless be possible
to address this problem in a suitable limit of the theory.

One of the great successes of General relativity was to replace the
gravitational force acting in flat space by motion of a free particle in
curved space along geodesics, incorporating in this way gravitation in the
principle of inertia. In the discussion that follows we will use $x^\mu$ to
describe either the evolution in time of a particle in space, $x^i(t)$, or
the evolution in proper-time of a particle in space-time, $x^\mu(s)$, and
$\dot x^\mu$ to represent the derivative with respect to its parameter. The
two cases differ mathematically only by the signature of the metric, which is
not relevant for the discussion. The geodesic equation is
\be
\ddot x^\mu+g^{\mu\nu}\left( \partial_\alpha g_{\beta\nu}-\frac{1}{2}
\partial_\nu g_{\alpha\beta}\right)\dot x^\alpha\dot x^\beta =0\quad ,
\label{geo}
\ee
and it defines the lines of shortest length computed using the given metric.

This equation can also be given another geometric interpretation, as the
autoparallels in a space with a rule for parallel transport, that is where a
linear connection $\Gamma_{\mu\nu}{^\chi}$ is defined. The autoparallels are
the lines along which the covariant derivative of the velocity vanishes,
\be
\ddot x^\mu+\Gamma_{\alpha\beta}{^\mu}\dot x^\alpha\dot x^\beta =0\quad .
\label{aut}
\ee
And this equation is equivalent to \re{geo} if one takes the connection to be
the Riemannian one, which is given in a coordinate system by the Christoffel
symbols,
\be
\bar\Gamma_{\alpha\beta}{^\mu}=\cri{\mu}{\alpha\beta}\ =\frac{1}{2}g^{\mu\nu}
(\partial_\alpha g_{\beta\nu}+\partial_\beta g_{\alpha\nu}-\partial_\nu
g_{\alpha\beta})\quad .\label{rcon}
\ee
The Riemannian connection is the only connection that can be
constructed solely from the metric tensor. Actually since the totally
antisymmetric Levi-civita symbol $\epsilon_{\alpha_1...\alpha_n}$ is
defined in any orientable manifold, for the particular case of three dimensions
the tensor $\sqrt{|g|}\epsilon_{\alpha\beta\gamma}g^{\gamma\mu}$ could be added
to the Riemannian connection in order to produce another connection. But such a
term would nevertheless
not affect the geodesic equation since it is antisymmetric in its first
two indices. And if we content ourselves with a metric description of space
(or space-time) the story would end here.

If a metric is not defined in space one can still have a definition of
parallel transport if a connection is given. In such a case only two tensors
can be constructed out of the connection (plus of course functions of
these),
\bea
\mbox{Curvature:}&&R_{\mu\nu\alpha}{^\beta}=\partial_\nu\Gamma_{\mu\alpha}
{^\beta}-\partial_\mu\Gamma_{\nu\alpha}{^\beta}+\Gamma_{\mu\alpha}{^\chi}
\Gamma_{\nu\chi}{^\beta}-\Gamma_{\nu\alpha}{^\chi}\Gamma_{\mu\chi}{^\beta}\\
\mbox{Torsion:}&&T_{\mu\nu}{^\beta}=\Gamma_{\mu\nu}{^\beta}-\Gamma_{\nu\mu}
{^\beta}\quad .
\eea
We note in passing that two natural and covariant conditions that can be
imposed on a space endowed with a connection are
\be
R_{\mu\nu\alpha}{^\beta}=0\label{rz}
\ee
or
\be
T_{\mu\nu}{^\beta}=0\quad .\label{tz}
\ee
Imposing both simultaneously would restrict the connection to be
trivial, that is, it would lead to ordinary flat space.

Since the work of Cartan, interest has been raised in looking upon space-time
as a space with parallel transport. But the metric is essential for the
description of Nature, so we should consider spaces with a connection and a
metric. Besides, also on physical grounds, one should impose the condition that
the metric is covariantly conserved, which is usually called the metric
condition. Doing so, the connection can be decomposed as
\be
\Gamma_{\alpha\beta}{^\mu}=\bar\Gamma_{\alpha\beta}{^\mu}+K_{\alpha\beta}{^\mu}
\quad ,\label{dec}
\ee
where the second term on the right hand side is a combination of torsion
tensors called the contorsion tensor,
\be
K_{\alpha\beta\gamma}=\frac{1}{2}(T_{\alpha\beta\gamma}-T_{\alpha\gamma\beta}-
T_{\beta\gamma\alpha})\quad .
\ee
This is the starting point for Einstein-Cartan theory where one considers
besides the metric the presence of torsion \cite{hel}. As we can see it differs
conceptually from Einstein's General relativity in looking upon space-time as
``a space with a connection and a metric that must be compatible among
themselves'' rather than as ``a metric space with a natural rule for parallel
transport derived from the metric''. Such theories gained
particular relevance after the analogy with the remaining interactions of
Nature (electroweak and strong) which are also formulated as theories of
connections and not as metric ones.

We remind the readers of the well known result that in a space with a metric
and a compatible connection, Riemannian geometry is recovered if the extra
condition of vanishing torsion \re{tz} is imposed, since then the connection
is expressible in terms of the metric by \re{rcon}. If we impose vanishing
curvature \re{rz} but not vanishing torsion, we are lead to teleparallel space,
which can be described with a set of tetrad fields by
\bea
g_{\mu\nu}&=&\vec e_\mu\cdot\vec e_\nu\\
\Gamma_{\mu\nu}{^\chi}&=&g^{\chi\omega}\vec e_\omega\cdot\partial_\mu\vec e_\nu
\quad .\label{nrcon}
\eea
The vector arrow refers to as many indices as the dimension of space and the
internal product is to be performed in those indices, using the flat metric
with the appropriate signature. Here and in what follows we will use
vector arrows and internal product dot notation in flat spaces, reserving the
indices notation for spaces where the metric is not trivial. We note that
both Riemannian and teleparallel spaces allow for arbitrary metrics. See for
example \cite{ricci} and references therein. See also Schr\"odinger
\cite{schr}.

In this context, of a space with parallel transport, the autoparallel equation,
\re{aut}, would seem more natural as a description of free particle motion.
And it must be noted that in such a space autoparallels and geodesics do not
coincide in general. But unlike in general relativity, since contorsion is a
tensor, it can be added freely to the autoparallel equation in order to
generate other covariant equations. In other words, it can be added or
subtracted to the original connection, generating other connections, so that
the connection is not unique anymore. In particular, the geodesic equation
shall still be a candidate. Actually, among the infinitely many candidates,
the most respectable ones are the autoparallels and the geodesics, for their
particular geometrical meanings, as the ``straightest'' lines and the
``shortest'' lines respectively.

The debate on whether geodesics or autoparallels should describe free particle
motion can however be carried on only as far as derivation of them from a
Lagrangean is not concerned. It has been known since long that from a
conventional variational principle only geodesics can be obtained \cite{hel}.

But recently Kleinert and collaborators proposed in a series of interesting
papers \cite{k2,k3,k4} a modification of the conventional variational
principle in torsionful spaces that leads to autoparallels as
the equations of motion for free particles.

\section{Tensor equations in non-coordinate frames}

Kleinert \cite{k2,k1} suggested that an inconsistency could be present in the
conventional way of deriving equations of motion in torsionful spaces by an
argument that goes as follows.

Starting from the Lagrangean for a free particle in flat space,
\be
L=\frac{1}{2}\dot{\vec x}\cdot \dot{\vec x}\quad ,\label{lag}
\ee
and performing a change of frame,
\be
\dot{\vec x}=\vec e_\mu\dot q^\mu\quad ,\label{tet}
\ee
 where the tetrads depend on the $q$-variables, $\vec e_\mu=\vec e_\mu(q)$,
one gets
\be
L=\frac{1}{2}g_{\mu\nu}\dot q^\mu\dot q^\nu\quad ,
\ee
with
\be
g_{\mu\nu}=\vec e_\mu\cdot \vec e_\nu
\ee
the induced metric on $q$-space.

Variation of this Lagrangean with respect to $q$ will lead as is well known to
the geodesic equation, \re{geo}. But variation with respect to $\vec x$ leads
to
\be
\ddot{\vec x}=0\quad ,
\ee
from which follows by the use of \re{tet}
\be
\ddot q^\mu+g^{\mu\nu}\vec e_\nu\cdot \partial_\alpha\vec e_\beta\dot q^\alpha
\dot q^\beta=0\quad .\label{aut2}
\ee
This is an autoparallel equation for the connection \re{nrcon}, which is in
general torsionful,
\be
T_{\alpha\beta}{^\mu}=g^{\mu\nu}\vec e_\nu\cdot (\partial_\alpha\vec e_\beta-
\partial_\beta\vec e_\alpha)\quad .\label{to}
\ee
The two results \re{geo} and \re{aut} are then in contradiction if the tetrads
do not obey $\partial_\alpha\vec e_\beta=\partial_\beta\vec e_\alpha$, that is,
if the transformation is not holonomic (integrable).

We would like to stress that this inconsistency is only apparent. This
disagreement between the two derivations is precisely what one should expect.
Paths in $x$-space and $q$-space are related by
\be
\vec x(t)=\int^t\vec e_\mu(q(s))\dot q^\mu(s)ds\quad .
\ee
If the transformation is non-holonomic a set of paths with fixed end points in
$x$-space is not mapped into a set of paths with fixed end points in $q$-space.
While the geodesic equation, \re{geo}, arises from a variation with fixed
endpoints in $q$-space, \re{aut2} arises from a variation with fixed endpoints
in $x$-space.

The conventional way of deriving equations of motion states that one should
vary coordinates. So, if $q$-space is a coordinate one, then the correct
equations of motion are the geodesics. Eq.(\ref{aut2}) is then wrong, for the
derivation of equations of motion in a non-coordinate frame ($x$-space in this
case) must follow a different procedure \cite{arn}.

On the other hand, if we assume $x$-space to be a coordinate one, then the
correct equations of motion are of course \re{aut2}. The connection \re{nrcon}
is not symmetric. However, since we are now assuming  $q$-space to be a
non-coordinate system, we have to account for the non-vanishing structure
coefficients
\be
c_{\alpha\beta}{^\mu}=g^{\mu\nu}\vec e_\nu\cdot (\partial_\alpha\vec e_\beta-
\partial_\beta\vec e_\alpha)\quad .\label{sc}
\ee
We further remind the reader of the generalization of the definitions of some
quantities to non-holonomic frames (cf. e.g. \cite{cb}):
\bea
R_{\mu\nu\alpha}{^\beta}&=&\partial_\nu\Gamma_{\mu\alpha}{^\beta}-\partial_\mu
\Gamma_{\nu\alpha}{^\beta}+\Gamma_{\mu\alpha}{^\chi}\Gamma_{\nu\chi}{^\beta}-
\Gamma_{\nu\alpha}{^\chi}\Gamma_{\mu\chi}{^\beta}+c_{\mu\nu}{^\chi}
\Gamma_{\chi\alpha}{^\beta}\\
T_{\mu\nu}{^\beta}&=&\Gamma_{\mu\nu}{^\beta}-\Gamma_{\nu\mu}{^\beta}-
c_{\mu\nu}{^\beta}\label{toa}\\
\bar\Gamma_{\mu\nu}{^\chi}&=&\cri{\chi}{\mu\nu}\ +\frac{1}{2}[c_{\mu\nu}{^\chi}
-g^{\alpha\chi}(g_{\beta\mu}c_{\nu\alpha}{^\beta}+g_{\beta\nu}
c_{\mu\alpha}{^\beta})]\quad .\label{nhs}
\eea

As can be seen from \re{nhs}, which involves besides the Christoffel symbols a
non-symmetric term, the Riemannian connection is in general not symmetric in a
non-coordinate frame, and it is indeed given in the present case by \re{nrcon}.
What we have called the torsion tensor, \re{to}, should then be interpreted as
the set of structure coefficients \re{sc}, and \re{aut2} represents in fact
the geodesic equation in this frame. We should also clarify that if torsion
vanishes in $x$-frame it will also vanish in $q$-frame, according to \re{toa}.
After all torsion is a tensor, if it vanishes in one frame it must vanish in
all frames.

The purpose of this section was to emphasize that there is no inconsistency in
the conventional variational principle, and to remind that the key point is
that variation with fixed endpoints in coordinate and non-coordinate systems
are not equivalent, and that a choice is needed, the conventional one being the
selection of coordinate systems, leading to the geodesic equation no matter
what the value of torsion is. We shall see in the next section that Kleinert
has proposed a different choice.

\section{The proposed variational principle}

Kleinert tried to obtain autoparallels from a variational principle by changing
the convention of varying the action with fixed end points in ``coordinate
systems'' to varying it in ``a non-coordinate system where the connection
vanishes''. This means that in the previous section, if $q$-space is a
coordinate system and the connection is given by \re{nrcon} in this system,
then the correct equations of motion are the ones derived in $x$-space, that is
\re{aut2}, which are the autoparallels. We shall use in the remaining of this
article this same meaning when we shall speak of $x$ and $q$ variables.

We shall not go into the details of the calculations, the interested reader may
look in the references \cite{k2,k3}. But it can be realized directly from the
description given above of the proposed variational principle that it has got a
restricted application because ``a frame where the connection vanishes'' can be
found if and only if the space is teleparallel, \re{rz}.

In his derivation \cite{k3} Kleinert starts from a Lagrangean that may depend
both on $q$-variables and on $x$-variables and arrives at a set of
integro-differential equations that we do not reproduce here and that are
non-causal. He reaches the conclusion that if and only if the Lagrangean
can be written solely in terms of $x$-variables,
\be
L=L\left(\dot{\vec x},\vec x\right) =
L\left(\vec e_\mu(q)\dot q^\mu,\vec x\right) \quad ,
\ee
then the equations of motion are causal, and they take the form of a system,
\bea
&&\vec e_\mu\cdot\frac{\partial L}{\partial\vec x}+\frac{\partial L}
{\partial q^\mu}-\dot{\left(\frac{\partial L}{\partial\dot q^\mu}\right) }-
T_{\mu\nu}{^\chi}\dot q^\nu\frac{\partial L}{\partial\dot q^\chi}=0\\
&&\vec x=\int^t\vec e_\mu(q(\tau ))\dot q^\mu(\tau )d\tau\quad .
\eea
We see thus that this derivation is even more restrictive than requirering
teleparallelism of space because it rules out the possibility of introducing
position dependent potentials. The remaining freedom of having a dependence
on $x$-variables will also be suppressed if we want the dynamical equations
to be differential and not integro-differential.

In a more recent paper \cite{k4} the previously described procedure was changed
by allowing $x$-space to have a higher dimensionality than $q$-space. The
vanishing curvature condition applies then to the whole of $x$-space, leaving
freedom for induced curvature in $q$-space, in much the same way as it happens
with the embedding of Riemannian manifolds in flat spaces by means of
holonomic mappings \cite{fri}.
We summarize this procedure by remembering that a torsionless $n$-dimensional
metric manifold can be described in terms of its embedding into a
$N$-dimensional flat manifold with $N>n$, and that a curvatureless
$n$-dimensional metric manifold can be described in terms of the embedding of
its tangent space
into a $n$-dimensional flat manifold. A general metric manifold with
non-vanishing curvature and torsion tensors could then be described by a
combination of the two methods, that is, the embedding of the tangent space
into a $N$-dimensional vector space with $N>n$.

This seems to be an interesting generalization of these methods and it also
seems to be the
correct formulation of Kleinert's original claims, though the authors left
still unproved that arbitrary curvature and torsion can be generated in this
way, and only a lower bound for the dimensionality of the embedding space was
computed. However it leaves unsolved the problem of the incorporation of
position dependent potentials.

We shall see in the next section that a bigger drawback affects this proposed
action principle even in its more recent form \cite{k4}: the lack of a
consistent hamiltonian formulation.

\section{Hamiltonian formulation and quantum\break mechanics}

In their paper \cite{k4} the authors commented on the fact that a hamiltonian
description was lacking. A Legendre transformation is always possible to
perform,
\bea
p_\mu&=&\frac{\partial L}{\partial\dot q^\mu}=g_{\mu\nu}\dot q^\nu\\
H&=&p_\mu\dot q^\mu-L=\frac{1}{2}g^{\mu\nu}p_\nu\quad ,\label{ham}
\eea
and phase space with a symplectic form can be defined in the usual way, along
with canonical transformations. Poisson brackets in canonical coordinates are
\be
\{ A,B\} =\frac{\partial A}{\partial q^\mu}\frac{\partial B}{\partial p_\mu}-
\frac{\partial B}{\partial q^\mu}\frac{\partial A}{\partial p_\mu}\quad .
\label{pab}
\ee

The difference with the conventional formulation comes with the time evolution.
Since in phase space we only need to specify the position variables endpoints
the variation of the action should follow the same procedure as in the
Lagrangean case, that is, varying in $x$-space and then coming back to
$q$-space. The hamiltonian formulation in $x$-space is trivial and leads to
\bea
H&=&\frac{1}{2}\vec\pi\cdot\vec\pi\\
\dot{\vec x}&=&\vec\pi\\ \label{x1}
\dot{\vec\pi}&=&\vec 0\quad .\label{p1}
\eea
Restricting ourselves to orbits in $x$-space that are mapped from $q$-space,
we have
\be
\vec\pi=\vec e_\mu g^{\mu\nu}p_\nu\quad.
\ee
From this equation and \re{tet} one gets
\bea
\dot q^\mu&=&g^{\mu\nu}p_\nu\\
\dot p_\mu&=&\Gamma_{\alpha\mu}{^\beta}g^{\alpha\gamma}p_\beta p_\gamma\quad .
\eea

This last set of equations describe correctly the autoparallel motion, as can
be checked by direct computation, and it can be derived from the hamiltonian
\re{ham}, the Poisson brackets \re{pab} and the time evolution equation
\be
\dot F=\{ F,H\} -T_{\mu\nu}{^\chi}\frac{\partial F}{\partial p_\mu}
\frac{\partial H}{\partial p_\nu}p_\chi\quad ,\label{hamil}
\ee
which will also give the right time evolution for any quantity defined in phase
space. However this time evolution is not generated by a hamiltonian vector
field. Consequently time evolution is not described anymore by a canonical
transformation and most of the classical results concerning hamiltonian systems
are not valid anymore. As an example, Liouville's theorem does not apply
anymore and the time derivative of the volume in phase space of an ensemble
of systems is given by
\be
\frac{dV}{dt}=\int_VT_{\mu\nu}{^\nu}\frac{\partial H}{\partial p_\mu}dv\quad .
\ee

We observe that one could also make the attempt of starting from the Lagrangean
\re{lag}, subject it to the constraints \re{tet} and follow Dirac's constrained
systems analysis. The constraints \re{tet} can actually be written without
reference to the embedded tangent space as
\be
\epsilon^{\mu_1...\mu_n}e_{\mu_1}{^{i_1}}(q)...e_{\mu_n}{^{i_n}}(q)
\epsilon_{i_1...i_N}=0\quad ,
\ee
being $n$ and $N$ the dimensions respectively of $q$ and $x$ spaces. But since
one cannot integrate the $q$, Dirac's method cannot be used.

The fact that the dynamics of this system is not hamiltonian is of course of
major importance. Concerning quantization that means that one cannot describe
time evolution by means of an unitary transformation and a prescription to
write the Schr\"odinger equation is missing. Kleinert \cite{k1} arrives to a
Schr\"odinger equation that we analyse in what follows.

\subsubsection*{Schr\"odinger equation}

Kleinert's extension of the Schr\"odinger equation can be obtained by replacing
in the equation
\be
i\partial_t\psi=-\frac{1}{2m}\nabla^2\psi
\ee
ordinary derivatives by covariant ones and contracting them with the metric.
We get
\be
i\partial_t\psi=-\frac{1}{2m}g^{ij}(\nabla_i\nabla_j-\Gamma_{ij}{^k}\nabla_k)
\psi\quad ,
\ee
which is equivalent to
\be
i\partial_t\psi=-\frac{1}{2m\sqrt{g}}\nabla_i(\sqrt{g}g^{ij}\nabla_j\psi)+
\frac{1}{2m}T^{ij}{_j}\nabla_i\psi\quad .\label{sch}
\ee

For a description of quantum mechanics in curved space see \cite{dew}. In a
coordinate representation by scalar wave functions, the inner product is
given by
\be
<\phi |\psi >=\int\sqrt{g}\phi^*\psi dx\quad ,
\ee
and the momenta conjugate to the coordinates are
\be
p_i=-i\left(\partial_i+\frac{1}{2}\bar\Gamma_{ij}{^j}\right)\quad ,
\ee
The hamiltonian corresponding to \re{sch} can then be written as
\be
H=\frac{1}{2m}g^{-1/4}p_ig^{1/2}g^{ij}p_jg^{-1/4}+\frac{i}{2m}T^{ij}{_j}g^{1/4}
p_ig^{-1/4}\quad .
\ee
The first term in this formula is obviously hermitian,
while the second, the one involving torsion, is not. Kleinert noticed
this problem and comments on it in sec.11.5 of his book \cite{k1}. However he
states that it can be solved for the particular application he has got in mind
(the hydrogen atom) and that in Nature it will not show up, invoking
Einstein-Cartan's theory where the torsion tensor is totally antisymmetric
\cite{hel}. But the point is that Einstein-Cartan's theory is precisely the
type of theory derived from a conventional variational principle that Kleinert
is questioning.

We should also notice that this extension of the Schr\"odinger equation cannot
ever lead to autoparallels in the classical limit for an arbitrary torsion,
because autoparallels depend on the symmetric part of torsion which has got
$n(n^2-1)/3$ independent components, and \re{sch} involves only the
contracted torsion tensor that has got $n$ independent components. Thus
\re{sch} cannot cover the general case of spaces with arbitrary metric and
torsion tensors.

\section{Final remarks}

\subsubsection*{Field theory}

All the considerations done so far in this article concerned ``particles'',
that is, the degrees of freedom were the coordinates themselves. But there is
the belief that matter should be described by fields. In this line of reasoning
the trajectories of free particles must be a suitable limit of a field theory.

But in a field theory this problem acquires a completely different form, for
the space coordinates become ``labels'' for the degrees of freedom and not the
degrees of freedom themselves. The proposed variational method that we have
been discussing does not make sense any longer for now the conventional
variational principle states that ``the values of the fields at any space point
must be fixed at the ends of the path''. Paths are not defined on space but
rather on the space of field configurations. Besides, using a non-coordinate
frame would affect the very counting of degrees of freedom, since the
transformation back to coordinates would act on what are now labels for the
degrees of freedom.

We are not imagining how to generalize the previous principle to fields. From
the point of view of the equations of motion themselves it seems that the
analogy would be to replace the Riemannian connection by the full connection
in the equations. The Klein-Gordon equation, for example, would become:
\bea
&g^{\mu\nu}D_\mu\partial_\nu\phi-m^2\phi=0&\Leftrightarrow\\
\Leftrightarrow&g^{\mu\nu}\partial_\mu\partial_\nu\phi-\Gamma_\mu{^{\mu\nu}}
\partial_\nu\phi-m^2\phi=0&\Leftrightarrow\\
\Leftrightarrow&\partial_\mu (\sqrt{-g}g^{\mu\nu}\partial_\nu\phi )-\sqrt{-g}
(T^{\nu\mu}{_\mu}\partial_\nu\phi +m^2\phi )=0&\quad ,
\eea
which is not derivable from a variational principle in the conventional way,
and which leads in the non-relativistic limit to Kleinert's extension of the
Schr\"odinger equation, \re{sch}.

We would like to note that it is possible to derive from a variational
principle a generalization of the Dirac equation to torsionful spaces
involving torsion, namely in Einstein-Cartan theory \cite{hel}. But in such an
equation only the totally antisymmetric part of torsion shows up. If one tries
the rule of replacing the ordinary derivative by the covariant one in the Dirac
equation, again non-hermitian terms will show up in the non-relativistic limit.
And such an equation will equally well be non-derivable from an action. The
symmetric part of torsion, precisely the one involved in the autoparallel
equation, is the one responsible for these undesirable features in field
equations.

\subsubsection*{Motivations for these ideas}

It has been stated several times that torsion arises naturally in a space with
parallel transport. That is true. Of course there is no reason why a connection
should be symmetric in a coordinate frame, as we saw in sec.1. But when dealing
with a metric space we also saw in that section that there is a naturally
defined connection, namely the Riemannian connection.
And insisting that the metric space should allow for the most general
connection compatible with it is by no means an obvious assumption.

Moreover vanishing of torsion provides the theory with the nice property that
antisymmetrized covariant derivatives are equal to ordinary exterior
derivatives on forms, essential ingredient in the construction of Lagrangians,
since for an arbitrary form $A_\mu$,
\be
D_\mu A_\nu -D_\nu A_\mu =\partial_\mu A_\nu -\partial_\nu A_\mu+
T_{\mu\nu}{^\chi}A_\chi\quad .
\ee

Our final remarks concern the original motivations for these works. They seem
to have originated from the description of defects in crystals by means of
torsionful spaces \cite{kro} (more recent works and applications in this field
can be found in \cite{k1,erik} and references therein). In this context
autoparallels would describe
the motion of an observer ``who would care only about the local crystalline
structure and follow his way as if he were in a defectless crystal''. This
description should be understood as an effective description of a much more
complicated system bearing no direct link to the fundamental principles of
mechanics. Moreover the actual framework for this theory of defects in
crystals seems to be best described in terms of teleparallel spaces rather than
in terms of spaces with arbitrary metric and torsion fields.

Another important motivation were the works of Kleinert on the quantization of
the hydrogen atom by the path integral method, using the Kustaanheimo-Stiefel
transformation \cite{k1}. It turns out however that both in this case and in
the case of the solid body equations of motion \cite{k5} the same results would
have been obtained in the conventional way by calling the frames used
non-holonomic ones and by calling torsion the structure coefficients. The
equations would be the same but ``autoparallels in a coordinate system in the
presence of torsion'' would be called ``geodesics in a non-holonomic frame, no
matter the value of torsion''. No example has been given in the quoted works of
a system with torsion that is not derivable from a torsionless space by a
non-holonomic transformation.

\section{Conclusions}

We conclude that the quoted works offer no decisive argument on why should one
change such fundamental principles of physics like the variational principle
and the result that geodesics describe the trajectories of test particles in
General relativity, and that on the contrary they raise new and difficult 
problems. The results of these works should probably be seen like useful tools
in effective theories where torsion may play a role. We
summarize the main disadvantages of the proposed variational principle:

- The variational principle proposed looks much more artificial than the
simpler principle of varying coordinates.

- It does not allow for the introduction of position dependent potentials.

- Time evolution is not described by a hamiltonian vector field in phase space
and consequently standard quantization breaks down.

- The Schr\"odinger equation proposed involves a non-hermitian hamiltonian and
fails to lead to autoparallels in the classical limit for an arbitrary torsion
field.

- The more fundamental problem of field theory falls out of this scheme.

- In a metric space the vanishing of torsion seems to be a natural and
convenient choice.

- The previous works that suggest the use of autoparallels are equally well
described by means of geodesics in non-holonomic frames.

- There is no experimental result suggesting the need for a modification of the
conventional variational principle.

\subsubsection*{Acknowledgements}

I thank Ingemar Bengtsson and Erik Aurell for discussions. This work was
supported by grant PRODEP-Ac\c c\~ao 5.2.


\begin{thebibliography}{99}
\bibitem{hel} F.W.Hell, P.van der Heyde, G.D.Kerlick, {\it General relativity
with spin and torsion: Foundations and prospects}, Rev.Mod.Phys. {\bf 48}, 393
(1976)
\bibitem{ricci} J.A.Schouten, {\it Ricci calculus}, Springer-Verlag (1954)
\bibitem{schr} E.Schr\"odinger, {\it Space-time structure},
Cambridge U.P.(1950)
\bibitem{k2} H.Kleinert, A.Pelster, {\it Lagrange Mechanics in Spaces with
Curvature and Torsion}, gr-qc/9605028
\bibitem{k3} P.Fiziev, H.Kleinert, {\it Anholonomic transformations of
Mechanical action principle}, gr-qc/9605046
\bibitem{k4} H.Kleinert, V.Shabanov, {\it Spaces with torsion from embedding
and the special role of autoparallel trajectories}, gr-qc/9709067
\bibitem{k1} H.Kleinert, {\it Path Integrals in Quantum Mechanics, Statistics
and Polymer Physics}, World Scientific (Singapore 1994)
\bibitem{arn} V.I.Arnold, V.V.Koslov, A.I.Neishtadt, in {\it Encyclopedia of
Mathematical Sciences, Dynamical Systems III, Mathematical Aspects of Classical
and Celestial Mechanics}, Springer-Verlag (Berlin 1988)
\bibitem{cb} Y.Choquet-Bruhat, C.DeWitt-Morette, M.Dillard-Bleick, {\it
Analysis, Manifolds and Physics}, North-Holland (Amsterdam 1977)
\bibitem{fri} A.Friedman, {\it Isometric embedding of Riemannian manifolds
into Euclidean spaces}, Rev.Mod.Phys. {\bf 37}, 201 (1965)
\bibitem{dew} B.S. DeWitt, {\it Point transformations in Quantum mechanics},
Phys.Rev. {\bf 85}, 653 (1952)
\bibitem{kro} E.Kr\"oner, {\it Continuum theory of defects} in R.Balian et al.,
{\it Physics of defects, Les Houches, Session XXXV, 1980}, 215, North-Holland
(1981)
\bibitem{erik} E.Aurell, {\it Torsion and electron motion in Quantum dots with
crystal lattice defects}, cond-mat/9711258
\bibitem{k5} P.Fiziev, H.Kleinert, {\it Motion of a rigid body in a body-fixed
coordinate system - For autoparallel trajectories in spaces with torsion},
hep-th/9503075

\end{thebibliography}
\end{document}